\documentclass[12pt]{article}

\usepackage{graphicx}

\title{Hadron form factors with the boost-corrected wave functions}

\author{    Yu.A.Simonov \\
 NRC ``Kurchatov Institute'' -- ITEP,\\ B. Cheremushkinskaya 25, \\Moscow, 117259, Russia}

\newcommand{\beq}{\begin{eqnarray}}
 \newcommand{\eeq}{\end{eqnarray}}
\newcommand{\be}{\begin{equation}}
 \newcommand{\ee}{\end{equation}}

\def\fun#1#2{\lower3.6pt\vbox{\baselineskip0pt\lineskip.9pt
\ialign{$\mathsurround=0pt#1\hfil ##\hfil$\crcr#2\crcr\sim\crcr}}}
\newcommand{\veX}{\mbox{\boldmath${\rm X}$}}
\newcommand{{\SD}}{\rm SD}

\newcommand{{\Mc}}{\mathcal{M}}
\newcommand{\veY}{\mbox{\boldmath${\rm Y}$}}
\newcommand{\vex}{\mbox{\boldmath${\rm x}$}}
\newcommand{\vey}{\mbox{\boldmath${\rm y}$}}
\newcommand{\ver}{\mbox{\boldmath${\rm r}$}}
\newcommand{\vesig}{\mbox{\boldmath${\rm \sigma}$}}

\newcommand{\veP}{\mbox{\boldmath${\rm P}$}}
\newcommand{\veA}{\mbox{\boldmath${\rm A}$}}
\newcommand{\vep}{\mbox{\boldmath${\rm p}$}}
\newcommand{\veq}{\mbox{\boldmath${\rm q}$}}
\newcommand{\veQ}{\mbox{\boldmath${\rm Q}$}}
\newcommand{\vez}{\mbox{\boldmath${\rm z}$}}

\newcommand{\veR}{\mbox{\boldmath${\rm R}$}}

\newcommand{\vek}{\mbox{\boldmath${\rm k}$}}

\newcommand{\vev}{\mbox{\boldmath${\rm v}$}}

\newcommand{\veB}{\mbox{\boldmath${\rm B}$}}

\newcommand{\veE}{\mbox{\boldmath${\rm E}$}}

\newcommand{\veal}{\mbox{\boldmath${\rm \alpha}$}}
\newcommand{\vepi}{\mbox{\boldmath${\rm \pi}$}}

\newcommand{\lan}{\langle}
\newcommand{\ran}{\rangle}

\begin{document}
\maketitle
\begin{abstract}
Hadron form factors are calculated using the Lorentz contracted wave functions, determined in the arbitrary dynamical scheme with the instantaneous interaction. It is shown that the large $Q$ asymptotics of the form factors is defined by the contraction coefficient $C_m(Q^2) = \frac{m}{\sqrt(m^2+ Q^2/4)}$, where $m$ is the meson mass, and weakly depends on the interaction used. The resulting form factors $F_\pi$ and $F_K$ are obtained in good agreement with the lattice and experimental data. Important consequences for the dynamics of hadron decays and scattering are shortly discussed.
\end{abstract}

 \section{Introduction}

The field theory is essentially the theory of the point-like objects, which interact via exchanges of point-like objects or in the external structureless fields. In this case  relativistic transformations are well known and bring about immediate results. However, in many cases one needs to describe the motion and interaction of extended objects and for that one has to know behavior of the Green's functions and the wave functions of extended objects under the applied boost, e.g. to know how the velocity $\vev$  of the system affects the hadron wave function.

As an example one can consider the form factor of a hadron, which can be defined in a relativistic  invariant way, but where the hadron wave function enters at two different momenta, $\vep_1$ and $\vep_2$. Another example is the hadron decay matrix element of the process $h\rightarrow h_1 + h_2$, e.g. $\rho\rightarrow \pi+\pi$, where the pions move with high velocity and therefore their wave functions enter in the strong decay matrix element in the Lorentz transformed way.

It is a purpose of present paper to derive the behavior of the hadron wave functions in the moving system and calculate the
resulting behavior of the meson form factors. As it is known, \cite{1} in the  relativistic  field theory the general formalism can be constructed in three different ways: 1/ the instant form, 2/ the point form and 3/ the light front form. In the instant form the wave function of any nonlocal object consisting of several elements can be defined at one moment of time and the frame (boost) dependence is dynamically generated in connection with Hamiltonian. In the literature different approaches have been developed for the practical realization of this problem, e.g. the quasipotential formalism in \cite{2*}, analysis of the operator matrix elements between wave functions and form factors \cite{3*},\cite{4*}. On another side the light front form was developed both for the 
 generalized parton distributions and form factors \cite{5*}.  As it is, the theory of the frame dependence of the Green's functions of any nonlocal objects  is closely related to the properties of the interaction terms in the Lagrangian, and one must envisage the instantaneous interaction for the first formalism, in particular confinement for the strong interaction and the Coulomb force in QED. The dynamical studies in this direction have been done
 recently, in Refs.~\cite{4,5,6} in several examples of systems. Later on, a more general and more phenomenological analysis was carried out in \cite{7}, where the properties of the spectrum and the wave functions in the moving system were studied in the framework of  the relativistic path integral formalism \cite{8,9,10,11}.  This method essentially exploits the universality and the Lorentz invariance of the Wilson-loop form of interaction, which produces both confinement and the gluon-exchange interaction in QCD. Moreover, in this formalism the Hamiltonian $H$ with the instantaneous interaction between quarks in QCD (called the relativistic string Hamiltonian (RSH)) and charged particles in QED was derived  and therefore the known defects of the Bethe-Salpeter approach are missing there. In \cite{7} it was shown that the eigenvalues and the wave functions, defined by the RSH, transform in the moving system in accordance with the Lorentz rules. Indeed, using the invariance law under the Lorenz transformations \cite{12,13},
\be
\rho(\vex,t)dV = {\rm invariant},~~\label{eq.01}
\ee
where $\rho(\vex,t)$ is the density, associated with the wave function $\psi_n(\vex,t)$,
\be
\rho_n(\vex,t) = \frac{1}{2i} \left(\psi_n \frac{\partial \psi_n^+}{\partial t}  - \psi_n^+ \frac{\partial \psi_n}{\partial t}\right) 
= E_n |\psi_n(\vex,t)|^2, ~~\label{eq.02}
\ee
and $dV=d\vex_{\bot} dx_{\|}$. One can use the standard transformations,
\be
L_{\rm P}dx_{\|} \rightarrow dx_{\|} \sqrt{1 - \vev^2}, ~~ L_{\rm P} E_n \rightarrow \frac{E_n}{\sqrt{1-\vev^2}},
\label{eq.03}
\ee
to insure the invariance of  (\ref{eq.01}). In its turn the invariance law implies that in the wave function $\psi(\vex,t)=\exp(-iE_nt)\varphi_n(\vex)$ the function $\varphi_n(\vex)$ is deformed in the moving system,
\be
L_{\rm P}\varphi_n(\vex_\bot, x_{\|}) = \varphi_n\left(\vex_\bot, \frac{x_{\|}}{\sqrt{1-\vev^2}}\right),
\label{eq.04}
\ee
and can be normalized as
\be
\int E_n |\varphi_n^{(v)}(\vex)|^2 dV_v = 1 = \int M_0^{(0)} |\varphi_n^{(0)}(\vex)|^2 dV_0,
\label{eq.05}
\ee
where the subscripts $(v)$ and $(0)$ refer to the moving and the rest frames. One of the immediate consequences from the
Eqs.~(\ref{eq.03}) and (\ref{eq.04}) is the property of the boosted Fourier component of the wave function:
\be
\varphi_n^{(v)}(\veq) = \int \varphi_n^{(v)}(\ver) \exp(i\veq\ver) d\ver = C_0 \varphi_n^{(0)}(\veq_\bot, q_{\|}\sqrt{1-v^2}),
\label{eq.06}
\ee
where $C_0 = \sqrt{1-v^2} = \frac{M_0}{\sqrt{M_0^2 + \veP^2}}$.

The equations (\ref{eq.01}) - (\ref{eq.06}) and in particular (\ref{eq.06}), formulated in Ref.~\cite{7}, will be the basic elements of our further analysis. In section 2 we shall write the expressions for the meson form factors in the Breit frame in terms of
the meson wave functions $\varphi_n^{(0)}$ in the rest frame.

The technic of the Fock-Feynman-Schwinger representation (FFSR) \cite{9,10,11} allows to represent the results in a simple form, which can be compared to experimental and lattice data in section 3. In section 4 we discuss the consequences and extrapolations of our results, as well as possible implications of the Lorentz contraction for the hadron decays and other processes. The concluding section contains a summary of results and discussion.

\section{Definition of the form factor through the $q\bar q$ Green's function}

As in Refs.~\cite{9,10}, we define the $q_1\bar q_2$ Green's function with the initial coordinates $\vex_1,\vex_2, t_4=0$ and the final coordinates $\vey_1,\vey_2,t_4=T$, which can be written as
\be
G(\vex_1,\vex_2,0|\vey_1,\vey_2,T) = \frac{T}{2\pi} \int^\infty_0 \frac{d\omega_1}{\omega_1^{3/2}}  \int^\infty_0
\frac{d\omega_2}{\omega_2^{3/2}} \left\lan Y_{\Gamma} \right\ran \left\lan \vex_1\vex_2| \exp(- H(\omega_1,\omega_2,\vep_1,\vep_2,) T)| \vep_1\vep_2 \right\ran,
\label{eq.07}
\ee
where $\left\lan Y_{\Gamma} \right\ran = tr \left\lan \Gamma (m_1- i\hat p_1) \Gamma (m_2 - i\hat p_2) \right\ran$, and in the presence of the electromagnetic field $A_{\mu}^{(e)}$ the Hamiltonian $H$ can be written as \cite{10,11},
\be
H = H_1(\omega_1) + H(\omega_2) + V_{12},
\label{eq.08}
\ee
with
\be
H_i(\omega_i) = \frac{(\vep_i - e_i \veA(\ver_i))^2}{2\omega_i}  + \frac{m_i^2 + \omega_i^2}{2\omega_i} + e_i A_0(\ver_i) -
 \frac{e_i (\vesig_i \veB)}{2\omega_i}  - i \frac{e_i (\veal_i\veE)}{2\omega_i},
 \label{eq.09}
 \ee
where $\veal_i = \left(\begin{array}{ll}0 & \vesig_i\\
\vesig_i & 0 \end{array}\right)$ and $V_{12}$ is
 \be
 V_{12} = V_0(\ver_1 - \ver_2) + V_{ss} + \Delta M_{se}.
\label{eq.10}
\ee
Here $V_0(r) = V_{conf}(r) + V_{oge}(r)$ and $V_{ss}$ is the spin-dependent part of the potential, while $\Delta M_{se}$ is the self-energy contribution to the mass, important for light and $s$ quarks \cite{11}.

The processes 1) $q_1\bar q_2 + \gamma \rightarrow q_1'\bar q_2$ and 2) $q_1\bar q_2 + \gamma \rightarrow q_1\bar q_2'$ can be described by perturbation due to the terms $e_iA_0$ or $\frac{e_i \vep_i\veA_i}{2\omega_i}$ of the $q_1\bar q_2$ Green's function with the initial total momentum $\veP_i$. In the case of conserving $\veP_i$ the equation (\ref{eq.07}) can be generalized,
$$
G_P(\vex_{12},0|\vey_{12},T) = \int d^3(\veX -\veY) \exp(i\veP(\veX - \veY)) \left\lan \veX,\vex_{12}| \exp(-H_P T)|\veY,\vey_{12}
\right\ran =$$ \be=\sum_n \varphi_P^{(n)}(\vex_{12}) \exp(- M_P^{(n)}(\omega_1,\omega_2) T) \varphi_P^{(n)+}(\vey_{12}),
\label{eq.11}
\ee
where the subscript $P$ implies the boosted eigenvalues and the eigenfunctions of the boosted Hamiltonian $H_P$.

At this point we need definitions of the c.m. and the relative coordinates,
\be
\veR = \frac{\omega_1\vex_1 + \omega_2\vex_2}{\omega_1 + \omega_2}, ~~\vex_{12} = \vex_1 - \vex_2,~~
\vex_1= \veR +\frac{\omega_2\vex_{12}}{\omega_1 + \omega_2},
\label{eq.12}\ee
and
\be
\veP = \vep_1 + \vep_2,~~\vek = \frac{1 \partial}{i \partial \vex_{12}}, ~\vex_2= \veR - \frac{\omega_1\vex_{12}}{\omega_1 + \omega_2}.
\ee
It is clear that $\omega_i$ are averaged in the integrations over $d\omega_i$ with the weight, shown in Eq.~(\ref{eq.07}),
yielding the stationary points $\omega_i^{(0)}=\left\lan \sqrt{m_i^2+\vep_i^2} \right\ran$  and, finally, the masses $M_{ik}$ as the eigenvalues of the Hamiltonian (\ref{eq.08}). Now expanding in $e_iA_0(\ver_i) = e_i\exp(i\veQ\ver_i)$, one obtains the first order form,
\be
\Delta G^{(1)} \equiv G(\vex_1,\vex_2;0|\vez_1,\vez_2;\tau) d^3z_1 d^3z_2 d\tau e_i A_0(z_i) G(\vez_1,\vez_2;\tau|\vey_1,\vey_2;T)
\label{eq.14}
\ee
Introducing the c.m. momenta $\veP,\veP'$ in the Green's function $G$ (\ref{eq.14}), as in Eq.~(\ref{eq.11}), one obtains
$$ \Delta G^{(1)} = \sum_{n,n'} \int d\Gamma_{\omega_1\omega_2} \varphi_P^{(n)}(\vex_{12}) \exp(- \bar{M}_P^{(n)}\tau) \varphi_P^{(n)+}(\vez_{12}) e_i d^3z_{12} \varphi_{P+Q}^{(n')}(\vez_{12})\times$$\be \times \exp(- \bar{M}_{P+Q}^{(n')}(T -\tau))
\exp\left(i \frac{\omega_2 \veQ\vez_{12}}{\omega_1 + \omega_2}\right) \varphi_{P+Q}^{(n')}(\vey_{12}).
\label{15}
\ee
From (\ref{15}) one can derive the $P-$dependent scalar form factor,
\be
F_P^{(n,n')}(Q^2) = \int \varphi_P^{(n)+}(\vez_{12}) \varphi_{P+Q}^{(n')}(\vez_{12}) \exp\left(i \frac{\omega_2 \veQ \vez_{12}}
{\omega_1 + \omega_2}\right) d^3z_{12}.
\label{16}
\ee
When the photon is absorbed by the quark $q_2$, then one has $A_0(\vez_2)=\exp (\veQ (\veR - \frac{\omega_1 \vez_{12}}{\omega_1 + \omega_2}))$ and the exponent in (\ref{eq.10}) becomes $\exp\left(-i\frac{\omega_1\veQ\vez_{12}}{\omega_1+\omega_2}\right)$. In what follows we define the universal factor, choosing $\veP = -\frac{\veQ}{2}$ in Eq.~(\ref{16}), as in the Breit frame; it yields the expression
\be
F^{(nn')}(Q^2) = \int \varphi_{-Q/2}^{(n)+}(\ver) \varphi_{Q/2}^{(n')}(\ver) \exp\left(i\frac{\omega_2\veQ\ver}{\omega_1 + \omega_2}\right)d^3r,
\label{eq.17}
\ee
where $\omega_i$ are taken at the stationary points, $\omega_i=\omega_i^{(0)}$.

\section{Boost dependent hadron wave functions}

In the quantum field theory (QED and QCD) the boost transformations of the wave functions were formulated in Eqs.~(\ref{eq.01}-\ref{eq.06}). The form factor in momentum space (\ref{eq.11}) can be written as
$$
F(Q^2) = \int \varphi_{-Q/2}(\vek)\varphi_{Q/2}\left(\vek+\veQ\frac{\omega_2}{\omega_1 + \omega_2}\right) d^3k =$$\be =
C_0^2(Q) \int \varphi_0(\vek_{\bot}, k_{\|}\sqrt{1-v^2}) \varphi_0\left(\vek_{\bot},\left(k_{\|} + Q \frac{\omega_2}{\omega_1+\omega_2}\right)
\sqrt{1-v^2}\right) \frac{d^3k}{(2\pi)^3}.
\label{eq.18}
\ee
Here $C_0(Q) = \frac{M_0}{\sqrt{M_0^2} + \frac{Q^2}{4}}$ and one can assume that in the c.m. system
the scalar, or the pseudoscalar, wave function is $\varphi_0(\vek) = \varphi_0(|\vek|) = \chi(\vek^2) = \chi(k_{\bot}^2 + k_{\|}^2)$. The Eq.~(\ref{eq.18}) is valid, when the momentum $\veq$ is given to the quark 1; in the case of the quark 2 one should replace the factor $\frac{\omega_2}{\omega_1+\omega_2} \rightarrow  \frac{\omega_1}{\omega_1 + \omega_2}$ and the total form factor is
\be
F_{\rm tot}(Q^2) = \frac{e_1}{e} F_1(Q^2) + \frac{e_2}{e} F_2(Q^2),
\label{eq.19}
\ee
and for the quarks $u,d,s,c$ one has $\frac{e_i}{e} = 2/3; -1/3;-1/3;2/3$. Note that $\omega_i$ are not proportional
to the quark or antiquark masses $m_i$ , instead in the confining string dynamics (\ref{eq.11}-\ref{15}) one has $\omega_i= \lan\sqrt{\vek^2 + m_i^2}\ran$ and e.g. in the $K^0$ meson the difference between $\omega_d$ and $\omega_{\bar s}$ is around 20\%, while in the $D$ and $D_s$ mesons this difference $\frac{\omega_{light}}{\omega_{heavy}}\cong (0.3-0.35)$ is not small.

In what follows, as the first approximation, we shall use the oscillator form of the wave function,
\be
\varphi_0(\vek) = \sqrt{\frac{8\pi^{3/2}}{k_0^3}} \exp \left(- \frac{k^2}{2k_0^2}\right),
\label{eq.20}
\ee
where $k_0$ is the only parameter of the hadron wave function. For hadrons the parameters $k_0$, corresponding to the wave functions, can be calculated in the FFSR \cite{15,16}; their values for different mesons are given in  Table~ (\ref{tab.01}), while for the $\pi$ and $K$ mesons they will be calculated later.

\begin{table}[!htb]
\caption{The parameter $k_0$ (in GeV) of the wave function (\ref{eq.20}) for different mesons}
\begin{center}
\label{tab.01}
\begin{tabular}{|l|c|}
\hline
 meson    &$ k_0 $ \\
\hline
 $\rho$   &  0.26 \\
 $D$      &  0.48 \\
 $B$      &  0.49 \\
 $\psi(2S)$ & 0.53 \\
 $J/\psi$   & 0.70 \\
 $\Upsilon$ & 1.27 \\
\hline
\end{tabular}
\end{center}
\end{table}
The accuracy of the approximation (\ref{eq.20}) can be shown to be $\sim 10\%$ for light mesons and around 1\% for heavy quarkonia.

Inserting (\ref{eq.20}) in (\ref{eq.18}) and integrating over $d^3k=d^2k_\bot dk_{\|}$, one finally obtains  the expression for the hadron form factor, coming from the first particle excitation,
\be
F_1(Q^2) = \frac{M_0}{\sqrt{M_0^2 + \frac{Q^2}{4}}} \exp\left(- \frac{Q^2 M_0^2 \nu_1^2}{4 k_0^2 (M_0^2 + \frac{Q^2}{4})}\right),
\label{eq.21}
\ee
where $\nu_1 = \frac{\omega_2}{\omega_1 + \omega_2}$. It is interesting to define the high $Q^2$ asymptotics at $Q^2\gg 4M_0^2$,
\be
F_1(Q^2\rightarrow \infty) = \frac{2M_0}{Q} \exp\left(- \frac{\nu_1^2 M_0^2}{k_0^2}\right).
\label{eq.22}
\ee
Note that due to boosting the wave functions in (\ref{eq.18}) never occurs in the high momentum region.
The behavior in Eq.~(\ref{eq.22}) agrees with the obtained in the first paper of \cite{4*} for spinless mesons,
where also instantaneous interaction between constituents was assumed.
\section{Lorentz contracted (pseudo)scalar meson form factors}

We start with the $\pi^+$ form factor. In this case, following (\ref{eq.18}), one can write
\be
F_{\pi}(Q^2) = C_{\pi} f_{\pi}(Q^2), ~~C_{\pi} = \sqrt{1 - v^2} = \frac{m_\pi}{\sqrt{m_\pi^2+\frac{Q^2}{4}}},
\label{eq.23}
\ee
where $f_\pi(Q^2)$ contains the pion wave function with different arguments,
\be
f_\pi(Q^2) = \int \frac{d^2q_\bot dk}{(2\pi)^3} \varphi_\pi(q_\bot, \kappa) \varphi_\pi,\left(q_\bot, \kappa+\frac{Q m_\pi}{2\sqrt{m_\pi^2 + Q^2/4}}\right).
\label{eq.24}
\ee
In Eq.~(\ref{eq.23}) one can see a remarkable feature, common to all (pseudo)scalar meson form factors -
at large $Q,~Q\gg m_\pi$, the dynamical part of the meson form factors - $f_\pi(Q^2)$ does not depend on $Q^2$ at all.
In the case of the pion this happens already for $Q^2\gg 4m_\pi^2$, or $Q\gg 0.28$~GeV.

One can call this phenomenon - the form factor freezing, which implies that the large $q^2$ asymptotics of the meson wave
function $\varphi(q^2)$ never defines the asymptotics of the form factor $F_\pi$ - contrary to the results of the perturbation
theory, predicting the $O(1/Q^2)$ behavior of the meson form factor at large $Q^2$. Of course, our result refers to the
main term of $F_\pi(Q^2)$, as in Eqs.~{(\ref{eq.23},\ref{eq.24}), and may not concern the correction terms, where the perturbation corrections are dominant.

Now we exploit the Gaussian form of the pion wave function (\ref{eq.20}) and obtain, as in Eq.~(\ref{eq.21}),
\be
f_\pi(Q^2) = \exp\left( - \frac{Q^2 m_\pi^2}{16 k_\pi^2 (m_\pi^2 + Q^2/4 )}\right).
\label{eq.25}
\ee
In what follows it is of interest to demonstrate the $Q^2$ behavior of  $C_\pi(Q^2)$ and  $f_\pi(Q^2)$  at different values of the Gaussian parameter $k_\pi$, presented in Table~\ref{tab.02}).
\begin{table}[!htb]
\caption{The function $C_\pi(Q^2)$  and the form factor $f_\pi(Q^2)$, as the functions of $Q^2$,  for different values of the Gaussian parameter $k_\pi$ (in GeV)}
\begin{center}
\label{tab.02}
\begin{tabular}{|l|c|c|c|c|c|c|c|}
\hline
$Q^2$ (in GeV$^2$)    & 0    &  0.35   &  0.60  &   0.75  &  1.0    &  1.60   &   2.45 \\

$C_\pi(Q^2)$           & 1.0    & 0.427   & 0.340  &  0.308 &  0.269  & 0.216  &  0.176  \\

$f_\pi(Q^2;k_\pi=0.20)$ & 1.0   & 0.903  &  0.895   & 0.892  &  0.890  &  0.887  &  0.886 \\

$f_\pi(Q^2;k_\pi=0.23)$ & 1.0   & 0.925  &  0.919   & 0.918   & 0.916  &  0.914   & 0.912 \\

$f_\pi(Q^2;k_\pi=0.25)$ &1.0    &  0.937  & 0.931   &  0.930 &   0.928  & 0.927  &  0.925 \\

\hline
\end{tabular}
\end{center}
\end{table}

From the $f_\pi(Q^2)$ values, given in Table~\ref{tab.02}, one can see that $f_\pi(Q^2) \cong 1.0$ with accuracy better that $10\%$ in the whole region $(0 - 2.45)$~GeV$^2$ and therefore the pion form factor $F_1(Q^2)$ can be represented by the
factor $C_\pi(Q^2)$ alone. Moreover, in the region $0.35\leq Q^2 \leq 2.45$~GeV$^2$  $f_\pi(Q^2)$ changes only within (1-2)\%, so that the main $Q^2$ dependence is defined by $C_\pi(Q^2)= \sqrt{1 - v^2}$.  The calculated here $F_\pi(Q^2)$ together with
experimental data on the pion form factor \cite{17,18,19} are given in Table~\ref{tab.03}.
\begin{table}[!htb]
\caption{Comparison of the experimental pion form factor \cite{19} with theoretical predictions from Eq.~(\ref{eq.23}-\ref{eq.25})}
\begin{center}
\label{tab.03}
\begin{tabular}{|l|c|c|c|c|c|}
\hline &&&&&\\
$Q^2$ (in GeV$^2$)            & 0.6         &    0.75           &  1.0          & 1.6         &  2.45\\\hline
&&&&&\\

$m_\pi$ (in GeV)                      &  0.20         &   0.191             &   0.181     &    0.168     &   0.166  \\\hline
&&&&&\\
$F_\pi(\exp.)$ [19]   & $0.433$ & 0.341 &0.312   &$ 0.233$ & $0.167$ \\

&$\pm 0.017$&$\pm 0.022$&$\pm 0.016$  &$ \pm 0.014$ & $\pm 0.010$\\\hline

&&&&&\\
$F_\pi(th.,k_\pi=0.25$~GeV)  &  0.316          & 0.286       &   0.25           &  0.20         &   0.163  \\\hline

&&&&&\\
$F_\pi(mod.,th.)  $        &    0.43     &     0.375          &   0.316         & 0.238         &    0.188 \\

\hline
\end{tabular}
\end{center}
\end{table}
In Table~\ref{tab.03}  the agreement between theoretical and experimental values within $O(25\%)$ accuracy takes place for all
$Q^2$ and with the accuracy better 10\% for $Q^2\geq 1.6$~GeV$^2$. This result was obtained for the simplest Gaussian form of the wave function and of course, can be improved with more suitable form of the wave function, which is important for
$Q^2\leq 1.0$~GeV$^2$. At the same time the asymptotic behavior, at  $Q^2\geq 1.6$~GeV$^2$, is given with a good accuracy and does not imply standard perturbative behavior \cite{20}. Note that at low $Q^2\leq 1.0 $~GeV$^2$ the agreement can be achieved using the dynamical $\pi$ meson mass and adding next terms of the oscillator basic expansion (see below).

We now turn to the $K^+$ meson form factor and as the first approximation, neglect the 20\% difference in the
values of $\omega_i$ of the $u$ and $s$ quarks and taking $\frac{\omega_1}{\omega_1+\omega_2}=1/2$. As a result, we come to the same Eqs.~(\ref{eq.23},\ref{eq.24}) and (\ref{eq.25}), where one should replace $m_\pi \rightarrow m_K$ everywhere; then for a rough estimate we take $m_K=0.5$~GeV and the momentum $k_K$ around the  value, calculated for the $\rho$ meson by solving the Hamiltonian equation (\ref{eq.09}), when $k_\rho=0.26$~GeV is obtained. In Table~\ref{tab.04} we give the values of the product $Q^2 F_K(Q^2)$   for $k_K=0.23$~GeV.
\begin{table}[!htb]
\caption{Comparison of the calculated function $Q^2F_K(Q^2)$ for the $K^+$ meson form factor ($k_K=0.23$~GeV) with experimental data \cite{21,23} and the lattice data \cite{24,25}}
\begin{center}
\label{tab.04}
\begin{tabular}{|l|c|c|c|c|c|}
\hline
$Q^2$                           & 0.10    &  0.5   &   1.0           & 1.5     &  2.5    \\

$Q^2F_K(Q^2)$ (th.)             & 0.0874  &  0.28  &  0.38           &  0.44   &  0.48  \\

$Q^2F_K(Q^2,exp.)$~\cite{22}     &        &        & $0.37\pm 0.12$  &         & $0.45\pm 0.04$ \\

$Q^2F_K(Q^2,lat.)$~\cite{24,25}    & 0.08     &  0.28  &  0.38          &          & 0.48  \\
\hline
\end{tabular}
\end{center}
\end{table}

From the numbers, given in Table~\ref{tab.04}, one can see that the function $Q^2F_K(Q^2)$, calculated here with several  approximations, occurs to be in very good agreement with the lattice results and experimental data.

At this point it is interesting to study the behavior of the $\pi$ and $K$ form factors at small $Q^2$, which values  are characterized by the effective radius, $F_{\pi,K}(Q^2) = 1 - \frac{Q^2}{6} r_{\pi,K}^2 + ...$. From experiment it is known that
$r_\pi^2(\exp.)= 0.44$~fm$^2$ \cite{20}, or $\frac{1}{6} r_\pi^2(\exp.) = 1.87$~GeV$^{-2}$, and
$r_K^2(\exp.)=0.34$~fm$^2$ \cite{22}, or $\frac{1}{6} r_K^2(\exp.)=1.445$~GeV$^{-2}$. At very small $Q^2$  in our theory one has the relation,
\be
\frac{r_{\pi,K}^2}{6} = \frac{1}{8 m_{\pi,K}^2} + \frac{1}{16 k_{\pi,K}^2}, ~~(Q^2\rightarrow 0).
\label{eq.26}
\ee
For the $K^+$ meson and chosen values $m_K=0.50$~GeV and $k_K=0.23$~GeV it gives $\frac{1}{6} r_K^2(th.)=1.68$~GeV$^{-2}$,
which is  16\% larger than experimental number. For the $\pi$ meson the situation is even worse and this fact shows that the
dynamics and the pion wave function (\ref{eq.23} -\ref{eq.25}) is not realistic at small $Q^2$. Indeed, the comparison of the experimental pion form factor at the values $Q^2=(0.35, 0.60, 0.75)$~GeV$^2$ with the theoretical numbers shows a (25-30)\% discrepancy. One can easily discover the root of it - the pion form factor up to $O(90\%)$ is determined by the pion mass and it is known  that the pion mass, described by the chiral theory, is shifted down from the dynamical value, $(0.35-0.40)$~GeV (as the singlet  partner of the $\rho$ meson), to the final value 0.14 GeV due to the GMOR relations \cite{26}.
The resulting pion Green's function, found in \cite{27}, has a very complicated form and is very different from the standard free Green's function with stable mass $m_\pi=0.14$ GeV. It can be expanded in a series of the Green's functions
with the masses, equal to non-chiral values of excited pseudoscalar states, starting with $m_1=0.35$ GeV.

We have exploited this fact, introducing the dynamical pion mass $m_1$, equal to 0.35~GeV at $Q^2=0$, and gradually decreasing it to the standard value 0.14 GeV at large $Q^2 g 2$~GeV$^2$, see $F_\pi(mod)$ in Table~\ref{tab.03}, where the resulting values of the pion form factor occur to be in good agreement with the experiment. Correspondingly, from Eq.~(\ref{eq.26}) and $m_\pi(Q^2)=0.35$~GeV one obtains $r_\pi^2/6(th.)=2$~GeV$^{-2}$ in good agreement with the experimental value,  $r_\pi^2/6=1.87$~GeV${-2}$ .

\section{Discussion and conclusions}

Our results, presented in Tables~\ref{tab.03} and \ref{tab.04}, show that suggested here formalism, which takes into account the Lorentz contraction of the meson wave functions, works reasonably well, especially for the $K^+$ meson, when with the only parameter - the Gaussian momentum $k_K$ - the experimental form factor is described quite well. However, the main idea of our method is not the reproduction of the experimental and lattice data in this case, but rather suggesting the theory, where the important physical phenomenon - the Lorentz contraction of the hadron wave function - defines the main features of the form factors of different hadrons and fully determines the character of its asymptotic behavior at large $Q^2$. The latter is given by the factor
$C_m (Q^2)= \sqrt{1 -v^2} = \frac{m}{\sqrt{m^2 + Q^2/4}}$, so that one expects the behavior
\be
F_m(Q^2 \rightarrow \infty) = const. \frac{m}{\sqrt{ m^2 + Q^2/4}}\sim const. \frac{2m}{Q},
\label{eq.27}
\ee
where the constant is defined by the asymptotics of the function as in Eq.~(\ref{eq.24}), but is $Q$-independent.
Indeed, if we even exploit the Coulomb wave function, $\varphi(p)=const/(p^2+b^2)^2$ to get a perturbative-like
result for the form factor, this wave function will enter in the form factor expression (\ref{eq.18}) as
$\varphi((p+Q)C_m)$, where $C_m$ behaves as $Q^{-1}$,  and yields the constant behavior at large Q.

This behavior basically differs from that, expected in the perturbation theory \cite{20}, or in the monopole and the dipole
equations, used in the fitting procedure. We do not enter here in important and difficult discussion about approaches to the theory of the hadron form factors, but we insist that if the effect of the Lorentz contraction is indeed present in nature, as we argue in our paper, then it should work in all physical theories, trying to describe  relativistic hadrons and their interaction, and in particular, in the approaches to hadron form factors.  This means that the effect of the Lorentz contraction of the wave function is more serious and should be included in any attempt to describe hadron form factor.

It is clear that the effect of the Lorentz contraction of the wave functions,
($(L_P\psi(\vex) = \psi\left(\vex_\bot,\frac{x_{\|}}{\sqrt{1 -v^2}}\right)$), can be present in many areas of physics and here we quote only a few of them: 1) the time-like form factors of hadrons and nuclei, where one finds the same high Q behavior as in the space-like form factors; 2) hadron decay amplitudes; 3) large $Q^2$ transfer reactions between hadrons or nuclei. Thus the Lorentz contraction effect opens new directions of developments, which can be used in the future.

The author is very grateful to A. M. Badalian for many discussions and important help in preparing this manuscript.

This work is supported by the Russian Science Foundation in the framework of the scientific project, Grant 16-12-10414.

\vspace{2cm}

{\bf Appendix A1.}  {\bf Boosted Hamiltonian and the boosted wave functions}  \\

 \setcounter{equation}{0} \def\theequation{A1.\arabic{equation}}

Here we follow the approach developed in Ref.~\cite{7,10,11}, where it was shown that the relativistic Hamiltonian can be written as $H_V = L(\veP) H_{c.m.}$,
\be
H_V = H_0 +L(\veP) V(\ver),~~ H_0 = \frac{\veP^2}{2(\omega_1 + \omega_2)} + \frac{\omega_1 + \omega_2}{2} +
\frac{\vepi^2}{2\tilde{\omega}} + \frac{m_1^2}{2\omega_1} +\frac{\omega_2^2}{2\omega_2}.
\label{A1.1}
\ee

Here $\omega_i$ are the virtual energies of the quarks 1 and 2 in the boosted system and $\vepi$ is the relative momentum in the same system. As shown in \cite{7}, the boosted instantaneous potential $V(\ver)$ , determined by the average Wilson loop,  and the wave function are defined as
\be
L(\veP) V(\ver) = \sqrt{1-v^2} V(\ver_\bot,r_{\|}\sqrt{1-v^2}); L(\veP)\Psi(\ver) = \Psi\left(\ver_\bot,\frac{r_{\|}}{\sqrt{1-v^2}}\right).
\label{A1.2}
\ee
In \cite{7} it was shown that the relative (virtual) momentum $\vepi$ can be expressed via the c.m. momentum $\vep$,
\be
\vepi =\vep;~ \pi_{\|} = p_{\|}\sqrt{1-v^2};~ \vep_\bot = \frac{1\partial}{i\partial \ver_\bot};~ p_{\|} =
\frac{1\partial}{i\partial r_{\|}}
\label{A1.3}
\ee
One can persuade oneself that the expressions (\ref{A1.2}) and (\ref{A1.3}) are internally consistent and the variable
$x_{\|}=\frac{r_{\|}}{\sqrt{1-v^2}}$ can be used in $\vepi^2$. Now we write the boosted virtual energy $\omega_i$ via its c.m. value
$\bar \omega_i$: $\omega_i=\frac{\bar{\omega}_i}{\sqrt{1-v^2}}$ and present $H_v$ as
\be
H_v\Psi = E_v\Psi; E_v = \frac{\veP^2}{2\Omega} + \frac{\Omega}{2} + \frac{M_0^2}{2\Omega},
\label{eq.A1.4}
\ee
where $\Omega=\frac{\bar \omega_1 + \bar \omega_2}{\sqrt{1-v^2}}$ and $M_0^2$ is defined via the eigenvalue of the
Hamiltonian equation in the c.m. system,
\be
(\frac{\vep^2}{2\bar \omega}+ \frac{m_1^2}{2\bar \omega_1} + \frac{m_2^2}{2\bar \omega_2} = V(\ver))\Psi_0 =
\frac{M_0^2}{2(\bar \omega_1 + \bar \omega_2)} \Psi_0,
\label{eq.A.5}
\ee
where the variable $\bar \omega= \frac{\bar{\omega}_1 \bar{\omega}_2}{\bar \omega_1 + \bar \omega_2}$, and $M_0^2$ is found,
taking the extremum of $M_0^2(\bar \omega_1,\bar \omega_2)$ in $\bar \omega_1, \bar \omega_2$ with the weight function, given in the path integral representation of \cite{10}, which yields $M_0^2(\bar{\omega}_1^{(0)}, \bar{\omega}_2^{(0)})$. Finally the extremum of $E_0(\Omega)$ gives $\Omega_0 = \sqrt{\vep^2 + M_0^2} = E_0(\Omega_0)$. This procedure was checked in \cite{7} in the case of the Coulomb and linear interactions.

\end{document}